\documentclass{kapproc} 

\usepackage{procps} 

\usepackage[dvips]{graphicx}

\upperandlowercase

\kluwerbib 

\begin{document}

\articletitle[Barred Galaxies] {Using Bars As Signposts of Galaxy
  Evolution at High and Low Redshifts}

\author{Kartik Sheth\altaffilmark{1}, Karin Menendez-Delmestre\altaffilmark{1}, Nick Scoville\altaffilmark{1}, Tom
  Jarett\altaffilmark{1}, Linda Strubbe \altaffilmark{1,2},Michael
  W. Regan\altaffilmark{3}, Eva Schinnerer\altaffilmark{4}, and David
  Block\altaffilmark{5}}
 
\affil{\altaffilmark{1}California Institute of Technology, Pasadena,
  CA, \altaffilmark{2}University of California - Berkeley,
  \altaffilmark{3}Space Telescope Science Institute,
  \altaffilmark{4}National Radio Astronomy Observatory,
  \altaffilmark{5}University of the Witwatersrand }

\begin{abstract}

An analysis of the NICMOS Deep Field shows that there is no evidence
of a decline in the bar fraction beyond z$\sim$0.7, as previously
claimed; both bandshifting and spatial resolution must be taken into
account when evaluating the evolution of the bar fraction.  Two main
caveats of this study were a lack of a proper comparison sample at low
redshifts and a larger number of galaxies at high redshifts.  We
address these caveats using two new studies.  For a proper local
sample, we have analyzed 134 spirals in the near-infrared using 2MASS
(main results presented by Menendez-Delmestre in this volume) which
serves as an ideal anchor for the low-redshift Universe.  In addition
to measuring the mean bar properties, we find that bar size is
correlated with galaxy size and brightness, but the bar ellipticity is
not correlated with these galaxy properties.  The bar length is not
correlated with the bar ellipticity.  For larger high redshift samples
we analyze the bar fraction from the 2-square degree COSMOS ACS
survey. We find that the bar fraction at z$\sim$0.7 is $\sim$50\%,
consistent with our earlier finding of no decline in bar fraction at
high redshifts.
\end{abstract}

\section{Bars: Signposts of Galaxy Evolution} 

Bars are ubiquitous in local disk galaxies (e.g., RC3, Eskridge et
al. 2000, Menendez-Delmestre et al. 2004).  They play an important
role in evolving galaxies by transporting vast amounts of gas to the
center (Sakamoto et al. 1997; Sheth et al. 2004), igniting
circumnuclear starburst activity (Ho et al. 1997 and references
therein), reducing the chemical abundance gradient (Martin \& Roy
1997), and perhaps feeding black holes (Shlosman et al. 1988, see
Knapen and Laurikainen in this volume).  Bars also form bulges, and
perhaps evolve galaxies along the Hubble sequence (e.g., Norman,
Sellwood \& Hasan 1996; see also Kormendy, Bournaud, and Combes in
these proceedings).  At high redshifts bars were expected to be fairly
common because of dynamically colder disks and increased merging
activity.  So it was a surprise when the first studies of high
redshift galaxy morphology found an apparent paucity of barred spirals
beyond z$\sim$0.5--0.7 (van den Bergh et al. 1996; Abraham et
al. 1999; van den Bergh et al. 2000).  If true these results set
strict constraints on cosmological simulations of galaxy formation by
requiring that disks were not sufficiently massive, or were
dynamically too hot until only $\sim$6 Gyr ago.  These constraints
would be in stark contrast to conclusions from studies of the cosmic
star formation history (e.g., Steidel et al. 1999 and references
therein) and thinness of local disks (e.g., Toth \& Ostriker 1992)
which argue that massive disks were already in place by z$\sim$1.

Bunker (1999) used a near-infrared 1.6$\mu$m NICMOS GTO image to show
that in at least one galaxy, a bar was missed by the early studies
because they observed the high redshift galaxies in the rest-frame
blue / ultraviolet light where bars are difficult to identify.  The
increasingly better visibility of bars at longer wavelengths is a
well-known effect (see for example Figure 1, and Menendez-Delmestre et
al and references therein in these proceedings).  We decided to
investigate whether the apparent decline in the bar fraction at high
redshifts could be due to such a selection effect?  This question was
the focal point of Sheth et al. (2003) and whose results we briefly
summarize here.

\begin{figure}[ht]
\centerline{\includegraphics[width=4in]{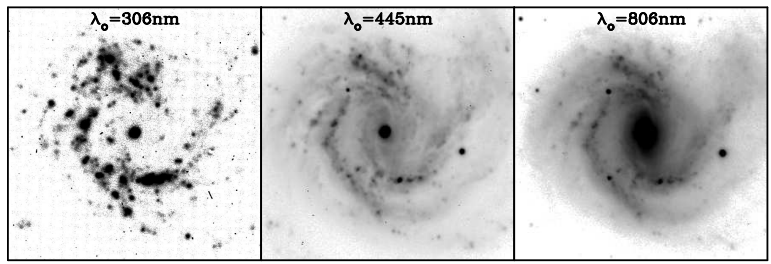}}
\caption{Left panel: UV appearance of NGC 4303, simulated using a
continuum-subtracted H$\alpha$ image.  Note how the bar is completely
invisible.  Middle panel: Blue-band image.  The bar is faint and
difficult to identify.  Right panel: I-band image. Only now does the
bar become visible.}
\end{figure}

\section{The Bar Fraction in the NICMOS Deep Field} 

We analyzed the bar fraction and bar properties in the WFPC2 (V, and
I-band) and NICMOS (H-band data) for the northern Hubble Deep Field
(Williams et al. 1996; Dickinson et al. 2000).  The NICMOS data are
ideal for studying galaxies in the redshift range 0.7$,$z$<$1 because
these data observe the galaxies in rest-frame V through I-band light.  

\subsection{The Bar Identification Technique} 

In contrast to previous studies which identified bars by eye, by
ellipticity breaks, or a change in isophote position angles between an
inner and outermost isophote (e.g., van den Bergh 1996; Abraham et
al. 1999;), we applied a more stringent two signature criteria to
identify bars.  We demanded that a) the ellipticity increase
monotonically and then drop with a sharp change of at least $\Delta
\epsilon$ > 0.1, and b) the position angle remain constant over the
bar region and change by $\Delta$ PA > 10$^\circ$ after the bar
region.  We require that the bar be symmetric by fixing the center for
the galaxy fitting algorithm.  Our algorithm misses bars if the galaxy
is highly inclined, if the bar position angle is the same as the
galactic disk, if the underlying galactic disk is too faint to be
adequately imaged, or if the data have inadequate resolution to
resolve bars.  Our method, though perhaps overly strict, has the
advantage of ensuring a firm lower limit to the bar fraction. 

As discussed in Sheth et al. (2003), at z $<$ 0.7, we identify five
barred spirals, and two candidate barred spirals, consistent with the
prior analysis of the HDFN by Abraham et al. (1999).  At z $>$ 0.7 we
identify four barred spirals and five candidate barred spirals,
including two possible candidates at z=1.66 and z=2.37.  The four
barred spirals are shown in Figure 2.  For comparison, in the previous
WFPC2 HDFN studies, van den Bergh et al. (1996) found no barred
spirals, and Abraham et al. (1999) found two barred galaxies beyond
z$\sim$0.5.  If we use the Abraham et al. (1999) magnitude cutoff of
I(AB)=23.7, the total number of disk-like galaxies drops to 31;
amongst these we identify three barred spirals.  Though we detect a
few more bars, the total number of barred spirals is still small.
Does this reflect a true decline in the bar fraction at z $>$ 0.7?  We
argue that when the spatial resolution of the observations are
considered in the context of bar visibility, there is no evidence of a
decline in bars at z$>$0.7.

\begin{figure}[ht]
\centerline{\includegraphics[height=4in]{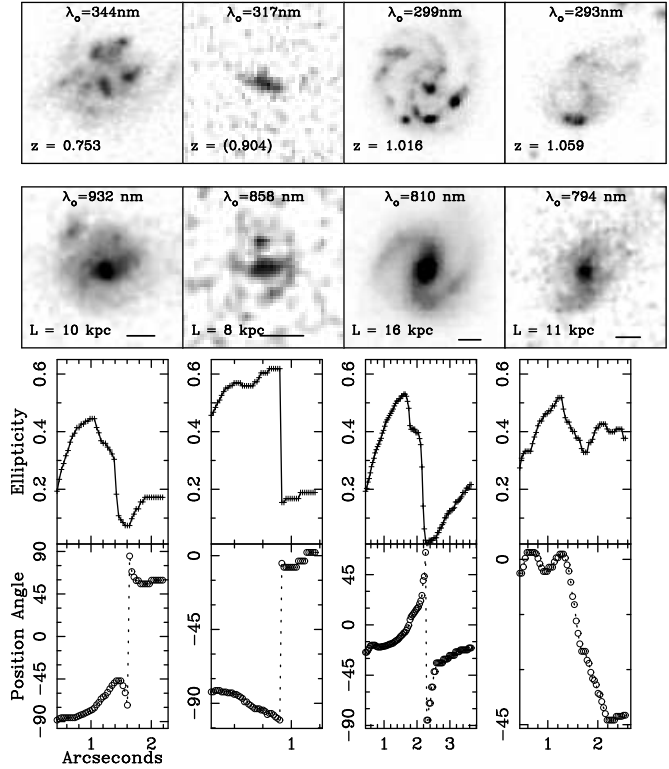}}
\caption{Barred spirals at z $>$ 0.7, arranged by redshift.  The top
row shows the optical (F606W, V-band) WFPC2 images and the second row
shows the near-infrared (1.6$\mu$m, H-band) NICMOS images.
0.5'' scale is shown with a horizontal segment in the lower
right of each panel.  The rest-frame wavelength for each galaxy is
listed inside the top of each panel.}
\end{figure}

\subsection{Spatial Resolution \& the Visibility of Bars}\label{resol}

In Figure 3, we show the apparent angular size of various galactic
structures as a function of redshift.  Overlaid are detection limits
for various telescopes adopting a five PSF detection threshold; five
PSFs is an appropriate choice (Menendez-Delmestre et al. 2004, also in
this volume).  The figure shows that at 0.8$\mu$m WFPC2 data is only
marginally capable of detecting a 5 kpc structure beyond z$\sim$0.7.
The NICMOS data have even coarser resolution (longer $\lambda$ and
larger pixels), and even though these data are not affected by
bandshifting until z $>$ 2--3, they only detect structures with sizes
$_>\atop{^\sim}$ 10 kpc.  The average size of the four bars identified
at z $>$ 0.7 is 12 kpc.

\begin{figure}[ht]
\centerline{\includegraphics[height=3in]{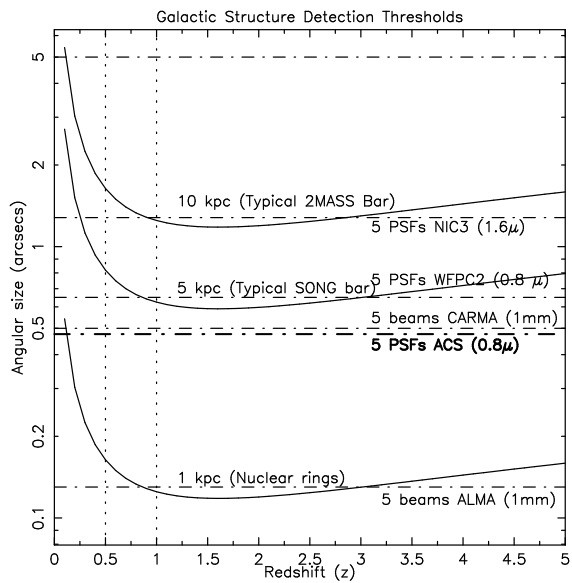}}
\caption{The detection threshold of various galactic structures as a
function of redshift for different telescopes and instruments is
shown.  The horizontal dotted-dashed lines are an arbitrary 5 PSF or 5
beam limit.  Note that at 0.8$\mu$m, even the WFPC2 data is only
marginally capable of detecting a typical 5 kpc bar at z $>$ 0.7; ACS
z-band is only slightly better.  The NICMOS data can only detect the
large bars at z $>$ 0.5.  Also shown are capabilities of two new
millimeter arrays, CARMA and ALMA which will be ideal for probing the
gas kinematics in high redshift systems.}
\end{figure}

The most important point of note from Figure 3 is that a measurement
of the bar fraction must take into account the size of bars and the
available spatial resolution of the data.  Although we only detect
three or four bars in the NICMOS data one must note that these data
are biased towards detecting only the largest bars.  Therefore, when
we compare the bar fraction at different redshifts, we must compare
the fraction for bars of equal sizes.  For a representative sample of
local galaxies (SONG, Regan et al. 2001, Sheth et al. 2004), we find
that bars with sizes $>$ 12 kpc are rare; only one out of 44 galaxies
in SONG has a bar larger than 12 kpc.  Thus the fraction of bars we
detected in the NICMOS Deep Field (4/95 galaxies for the entire
sample, or 3/31 galaxies using an I-magnitude cutoff) is similar, and
perhaps even, larger than the bar fraction seen in SONG.  Thus we
concluded that there was no evidence for a decline in the bar fraction
at z $>$ 0.7, as previously claimed.

Our results are hampered by small number statistics (we discovered
only a handful of bars at z > 0.7 and our local comparison sample
contained forty four galaxies), and difficulty in defining comparable
samples at high and low redshifts (see Sheth et al. 2003).  From the
existing analysis, it is difficult to conclusively state how the bar
fraction varies as a function of redshift, and how the bar properties
vary.  In the next two sections we outline new results from two
separate on-going studies aimed at overcoming these two main caveats
of our NICMOS study.

\section{Defining a Proper Local Sample: The 2MASS Local Galaxy Atlas} 

How well do we know the fraction of nearby bars?  What are their
properties (size, strength)?  How do these properties depend on the
host galaxy properties?  Answers to these questions are of fundamental
importance before any comparison is made to the high redshift
Universe.  

Since bars are best studied in the infrared, we decided to address all
of these questions using the 2MASS Large Galaxy Sample.  Results of
this study were presented in a poster at this conference by
Menendez-Delmestre.  We chose all spirals of type Sa-Sd, with
$i<$65$^0$ for a sample of 134 galaxies in J+H+K.  We ran the same
ellipse fitting algorithm on these galaxies as we did previously on
the high redshift sample and identified bars using the two signature
criteria described above.

As shown in Figure 2 of Menendez-Delmestre et al. (this volume), the
fraction of barred spirals in the 2MASS sample is 58\% (see Figure 1
in Menendez-Delmestre et al.).  Another 21\% of the galaxies are
identified as candidate bars where usually there is an ellipticity
signature but no corresponding position angle change.  We examined
each of the candidate bars by eye and found that about two-thirds of
the candidates were in fact barred.  So the total fraction of barred
spirals in the infrared $\sim$72\%, similar to the fraction found by
Eskridge et al. (2000).  The bar fraction changes slightly with the
T-type with the highest fraction (80\%) in T=3 but there is no
significant trend in the fraction with the Hubble type.  Unfortunately
the 2MASS sample has low signal to noise and we were unable to
quantify the bar fraction in galaxies of type later than Sd.  Note
that the overall bar fraction in RC3 is lower, but not significantly
different than the 2MASS fraction.  This suggests that overall the
change in the bar fraction between B-band (RC3) and the near-infrared
is not large but also note that the effect may become severe at
wavelengths shorter than the B-band, as shown by Figure 1.

\begin{figure}[ht]

\centerline{\includegraphics[width=4in]{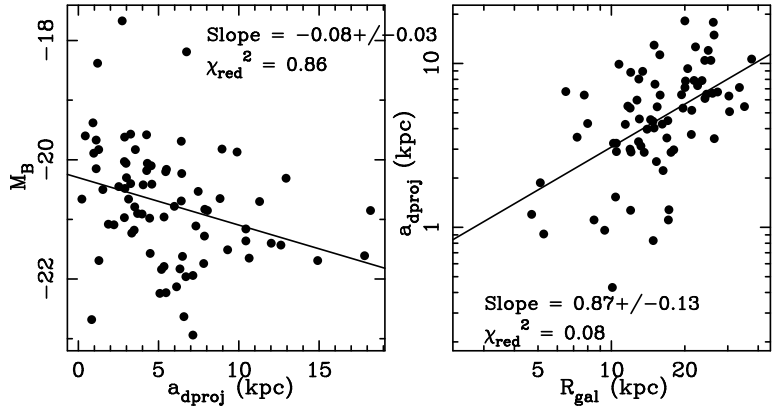}}
\caption{Left Panel: Absolute B-magnitude vs. deprojected bar
  semi-major axis.  Right Panel: Deprojected bar semi-major axis
  vs. galaxy size.  There are weak trends of larger bars in brighter
  and larger galaxies.  }

\end{figure}
For the 78 barred spirals in 2MASS, we derive a median deprojected bar
semi-major axis of 5.1 kpc.  This result is different from the SONG
sample we considered in Sheth et al. (2003).  The reason for the
difference is that 2MASS surveys a larger volume but is shallower.  As
a result we observe larger and brighter galaxies which tend to have
longer bars (see below).  We measure a mean ellipticity of 0.45 but
the most notable result from the distribution of ellipticities is the
scarcity of thin bars ($\epsilon > 0.7$).  Previously similar results
have been interpreted as evidence for a second or later generation of
bars (e.g., Block et al. 2002, see also Bournaud, Combes in this
volume) because in numerical simulations the first generation of bars
is expected to be strong (thin) whereas subsequent generations are
expected to be weaker (fatter).  However note that the measurement of
ellipticity can be affected by bulge light.  Ellipse fitting does not
automatically account for bulge light and as a consequence gives a
higher ellipticity than the actual value.  When we compare the bar
ellipticity and bar length we find no correlation indicating that
larger bars are preferentially fatter or thinner.  We find that bars
come in all shapes and sizes.

How do the bar properties depend on the host galaxy?  From the 2MASS
data, we find a trend of larger bars in larger and brighter galaxies
(Figures 4).  This is expected given that the bar instability is
correlated with the mass of the disk.  

Correlations of the bar strength (we use ellipticity as a proxy for
strength) and the host galaxy parameters are very weak (Figures 5).
Unlike the bar length, there is no correlation between the bar
ellipticity and galaxy size or brightness.

\begin{figure}[ht]

\centerline{\includegraphics[width=4in]{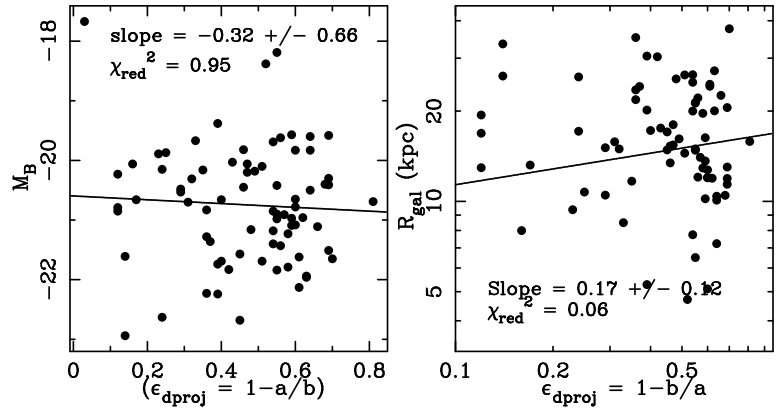}}
\caption{Left Panel: Absolute B-magnitude vs. deprojected bar
  ellipticity.  Right Panel: Size of galaxy vs. deprojected bar
  ellipticity.  There is no correlation between the bar ellipticity
  and the host galaxy size or brightness.  }

\end{figure}

Note that these correlations are not immune to biases in the analysis.
For instance the ellipse fitting technique always underestimates the
bar ellipticity (Sheth et al. 2000, 2004) because the bulge light
affects the galaxy isophotes to make the fitted ellipses
fatter/thicker than the real underlying bar.  Hence the lack of
``strong'' bars in early-Hubble type galaxies is not necessarily a
real effect.  Another bias is related to resolution.  Small bars are
difficult to identify especially in galaxies with bright bulges and
nuclei.  So any trend of larger bars in earlier type galaxies should
also be viewed with caution.  Nevertheless the trends of larger bars
in larger galaxies and brighter galaxies are relatively robust.

\section{COSMOS:  Overcoming Small Number Statistics}

The 2MASS analysis by Menendez-Delmestre et al. (2004) offers an
excellent analysis of the local sample of bars and their host
galaxies.  We can now extend these results to higher redshifts to
understand how the bar fraction and bars evolved over time.

\begin{figure}[ht]
\vskip-0.2in
\centerline{\includegraphics[height=3in]{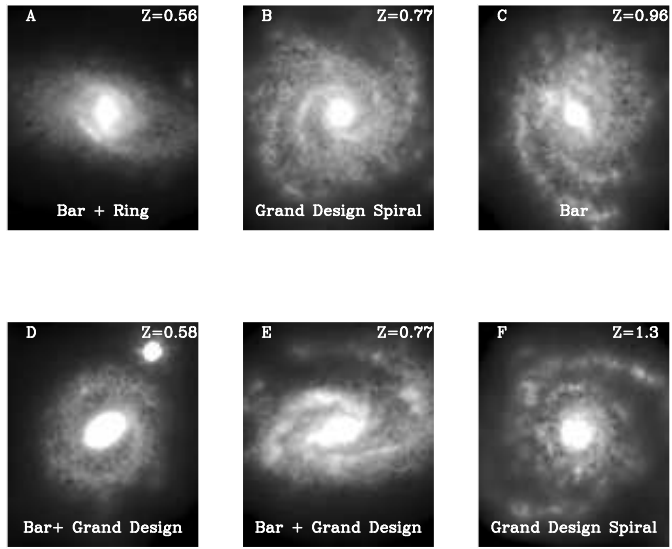}}
\caption{Typical barred and spiral galaxies from COSMOS data.  Galaxy
  labeled A is virtually a twin of NGC 4303 shown in Figure 1.  It
  shows a strong bar and a ring encircling the bar.  The redshifts are
  photometrically derived from ground-based BVRIz Subaru data and are
  reliable to z~0.8.  So galaxies C and F may in fact be at lower
  redshifts than indicated here.  }
\end{figure}
Until recently, the Hubble Deep Fields were the standard windows into
the high redshift Universe.  These data already indicated that
defining a comparative sample at high redshifts is difficult.  For
example the fraction of irregular objects increases to nearly 30\% to
z$\sim$1 (e.g., Griffiths et al. 1994, Abraham et al. 1996).  Lilly et
al. (1996) concluded that disks evolve passively in luminosity whereas
Shude et al. (1998) argue in favor of significant evolution.
Certainly some of these discrepancies are a result of cosmic variance
and scarcity of high quality data.  These gaps, however, are now being
filled with surveys like GOODS, UDF, DEEP2, and GEMS.

The most ambitious and revolutionary new survey is COSMOS
(http://www.astro.caltech.edu/cosmos) which is observing a contiguous
2-square degree equatorial field using the Advanced Camera for Survey
on the Hubble Space Telescope, and multi-wavelength data from X-rays
to radio. The size of the field is important because it overcomes
cosmic variance and provides a truly representative view of the high
redshift Universe.  COSMOS is an ideal treasury program with a vast
range of applications.  In particular, for the study of galaxy
morphology COSMOS is a gold mine (or more appropriately for this
conference, a diamond mine) of opportunity.  It will provide
spectroscopic redshifts for over 100,000 galaxies with 10,000 galaxies
in each of five bins of from z=0.5 to z=2.  The spatial resolution of
ACS data is exquisite.  As shown in Figure 3, one can easily identify
structures with sizes as small as 3 kpc in diameter. If Figure 6 we
show some typical examples of disk galaxies at three photometrically
determined redshifts.

\subsection{The Bar Fraction in COSMOS} 

\begin{figure}[ht]
\vskip-0.1in
\centerline{\includegraphics[width=3in]{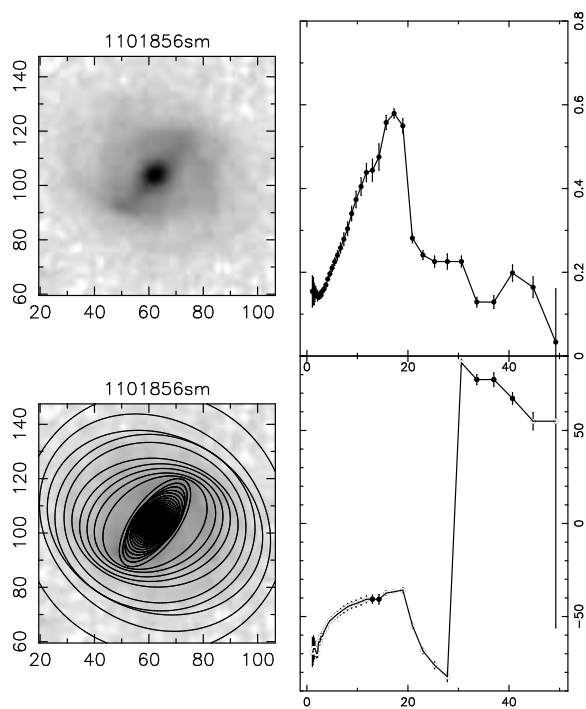}}
\caption{Example of ellipse fitting for a barred spiral in COSMOS.}

\end{figure}

In order to image such a large field with adequate sensitivity (COSMOS
is within half a magnitude of the GOODS data) we chose to use the
broader I-band filter over the z-band filter for COSMOS.  Hence the
analysis of the bar fraction can be done with full confidence to
z$\sim$0.7--0.8 and perhaps to even higher redshifts.  To date
approximately 200 fields out of 600 have been observed.  For these
fields we first identified a sample of spiral galaxies from a
photometric catalog derived from BVRIz data from observations on the
Subaru teelscope for the entire COSMOS field. We only chose those
galaxies that had photometric redshifts derived with the highest
confidence ($>$95\%).  We ran the same ellipse fitting algorithm on
these galaxies.  An example of such a fit is shown in Figure 7.  We
have done a preliminary analysis for the redshift bin 0.6 $ < $ z $ <
$0.8 for $\sim$ 700 galaxies.  We find that the fraction of barred
spirals is $\sim$50\%, consistent with the local fraction. Although
further analysis is necessary with better redshifts and the complete
sample, the results already indicate that the bar fraction is not
declining significantly, at least out to z$\sim$0.8.

\end{document}